\documentclass[11pt]{article}
\usepackage[utf8]{inputenc} 
\usepackage[T1]{fontenc}
\usepackage[a4paper,left=3cm,right=3cm,top=3cm,bottom=3cm]{geometry}
\usepackage[english]{babel} 
\usepackage{hhline}
\usepackage{amsmath} 
\usepackage{amsfonts}
\usepackage{amssymb}
\usepackage{graphicx}

\usepackage[bookmarksopen=true, hidelinks]{hyperref}
\usepackage{bookmark}
\usepackage[toc,page]{appendix}
\usepackage{url}
\usepackage{color}
\usepackage{shadow}
\usepackage{fancybox}
\usepackage[babel=true,autostyle=false, style=english]{csquotes}
\MakeOuterQuote{"} 

\usepackage{algorithm,algorithmicx}
\usepackage{algpseudocode}
\usepackage{caption}
\usepackage{parskip} 

\def\N{\mathbb{N}}
\def\R{\mathbb{R}}

\usepackage{authblk}
\author[$*$,1]{Alexis Coyette}
\author[$*$,1]{Charles Modera}
\author[$*$,1]{Candy Sonveaux}
\author[$*$,1]{Judica\"el Mohet}
\author[$*$,1]{François-Grégoire Bierwart}
\author[$*$,1]{Sylverio Pool Marquez}
\author[$*$,1]{Jarod Ketcha Kouakep}
\author[$*$,1]{Cédric Simal}
\author[$*$,1]{Komlan Fiagbe}
\author[$*$,1]{Violaine Piengeon}
\author[$*$,1]{Martin Moriamé}
\author[$*$,1]{Justine Bodart}
\author[$*$,1]{Marie Dorchain}
\author[$*$,1]{Maxime Lucas}
\author[$*$,1]{Rommel Tchinda Djeudjo}
\author[$*$,1,2]{Gianluca Peri}
\author[,1]{Eve Tilman\thanks{Unequal contribution. Author order determined by the proposed ranking method.}}
\affil[1]{Department of Mathematics, Naxys, Unamur, Belgium}
\affil[2]{Department of Physics and Astronomy, CSDC, University of Florence, Italy}

\date{April 1, 2026}

\title{On rankings in multiplayer games with an application to the game of Whist}

\begin{document}
\maketitle


\begin{abstract}
    We propose a novel extension of the Bradley-Terry model to multiplayer games and adapt a recent algorithm by Newman \cite{Newman2023} to our model. We demonstrate the use of our proposed method on synthetic datasets and on a real dataset of games of cards.
\end{abstract}

\section{Introduction}

The Bradley-Terry (BT) model is a popular statistical model for modelling the outcomes of pairwise comparisons between a finite number of items, such as tournaments in sports or games \cite{zermeloBerechnungTurnierErgebnisseAls1929,bradleyRankAnalysisIncomplete1952,fangRecentAdvancesBradleyTerry2026}, with recent applications in preference learning of Large Language Models \cite{Christiano2017,Ouyang2022}. 

The parameters of this model are latent "player strengths" which characterize the probability of victory of a given player over another, and enable using these parameters to establish rankings of players. The traditional algorithm for inferring these parameters via log-likelihood maximization goes all the way back to the seminal paper by Zermelo \cite{zermeloBerechnungTurnierErgebnisseAls1929}, and was recently improved upon by Newman \cite{Newman2023}.

The classical BT model assumes comparisons are made between pairs of individual players, or teams of players considered as a whole \cite{dae-kikangPoissonModelBradley2015}. There are, however, many situations where games involve more than two players, such as horse races, or tennis doubles. The first case, where the outcome of a single game is a ranking of all the involved players, is described by the Plackett-Luce model \cite{Luce1959,plackettAnalysisPermutations1975}, for which \cite{yeungEfficientInferenceRankings2025} recently adapted Newman's algorithm. In this work, we study the latter case, in which the outcome of a single game is a directed relation between two teams of players. Our contributions are the following

\begin{itemize}
    \item We extend the BT model to the setting of games between teams of players, which may change from game to game.
    \item We adapt Newman's algorithm to obtain an efficient method to estimate player strengths from a recorded set of games.
    \item We compare our method to the prior work of Huang et al. \cite{Huang2004}, for which we derive a new update rule based on our approach.
\end{itemize}

The rest of this note is structured as follows. In Section \ref{sec:methods}, we present our model along with the inference rule for its parameters. We then compare it with prior work in this area in Section \ref{sec:other-works}. Finally, we evaluate our model first on synthetic data, then on a real dataset in Section \ref{sec:experiments}.

Detailed derivations of the various formulas are delegated to the Appendix. In addition, we defer theoretical results such as proofs of convergence to future work.

\section{Methods} \label{sec:methods}

\subsection{The Bradley-Terry model}

The classical Bradley-Terry model, initially discussed by Zermelo \cite{zermeloBerechnungTurnierErgebnisseAls1929}, then later rediscovered \cite{bradleyRankAnalysisIncomplete1952}, models the outcome of competitive two-player games within a pool of $n\in \N$ players. We denote by $P[i \leftarrow j]$ the probability of player $i$ winning against player $j$, which is assumed to only depend on the latent "strength" of each player. The BT model, in particular, assumes the following form, 
\begin{equation} \label{eq:bt-prob}
    P[i \leftarrow j] = \frac{\pi_i}{\pi_i + \pi_j},
\end{equation}
where $\pi_i > 0$ denotes the skill of player $i \in \{1,\dots,n\}$. Equation \eqref{eq:bt-prob} can be shown to be equivalent to 
\begin{equation} \label{eq:bt-prob-2}
    P[i \leftarrow j] = f(s_i - s_j),
\end{equation}
where $s_i = \log \pi_i$ is the \textit{strength} of player $i$, and $f(s) = 1/(1 + e^{-s})$ is the logistic function. This last equation is invariant under adding an arbitrary constant to the variables $s_i$ (equivalently, equation \eqref{eq:bt-prob} is invariant under scaling the variables $\pi_i$ by an arbitrary positive constant).

Given a set of recorded games, represented as a matrix $W \in \R^{n \times n}$ such that $W_{ij}$ is the number of times player $i$ won against player $j$, approximate skills $\hat{\pi}_i$ can be estimated by maximizing the log-likelihood
\begin{equation} \label{eq:bt-loglikelihood}
    \log P(W | \pi) = \sum_{i < j} W_{ij} \log \left(\frac{\pi_i}{\pi_i + \pi_j}\right) + W_{ji} \log \left( \frac{\pi_j}{\pi_i + \pi_j} \right).
\end{equation}
Taking the gradient of this expression with respect to $\pi = [\pi_1, \dots, \pi_n]^\top \in \R^n_+$ and writing the first order optimality condition, one can arrive at the following set of fixed-point equations, known since \cite{zermeloBerechnungTurnierErgebnisseAls1929},
\begin{equation} \label{eq:bt-zermelo}
    \pi_i = \frac{\sum_{j} W_{ij}}{\sum_{j} (W_{ij} + W_{ji})/(\pi_i + \pi_j)}, \,\,\,i \in \{1,\dots,n\}.
\end{equation}
Iterating this equation gives a simple algorithm for inferring the values $\pi_i$, which is guaranteed to converge if $W_{ij}>0, \forall i,j$ (more generally, if we interpret $W$ as the adjacency matrix of a directed graph, the algorithm will converge if this graph is strongly connected \cite{zermeloBerechnungTurnierErgebnisseAls1929}). In practice, it is common to apply the update \eqref{eq:bt-zermelo} one parameter at a time, and to apply some form of normalization such as imposing $\sum_{i=1}^n \pi_i = 1$ or $\sqrt[n]{\prod_{i=1}^n \pi_i}=1$.

While the algorithm described above is guaranteed to converge to a global maximum of the log-likelihood under certain conditions, its convergence is known to be quite slow. Recent work by Newman proposed a simple modification of \eqref{eq:bt-zermelo} which converges much faster while having all the same guarantees \cite{Newman2023},
\begin{equation} \label{eq:bt:newman}
    \pi_i = \frac{\sum_j W_{ij} \frac{\pi_j}{\pi_i + \pi_j}}{\sum_j W_{ji} / (\pi_i + \pi_j)}.
\end{equation}

The BT model is not the only one of its kind, and there are many alternatives taking different functions for Equation \eqref{eq:bt-prob-2}, (see e.g. \cite{fangRecentAdvancesBradleyTerry2026} for a list of examples), allowing for ties, or accounting for the intrisic "randomness" of certain games \cite{jerdeeLuckSkillDepth2024}. Those models often do not have simple inference algorithms like the BT model, and must usually be fit using generic methods such as Bayesian inference, which are computationally more expensive.

\subsection{The BT model on directed hypergraphs}

The BT model is appropriate for establishing rankings in two-player games, and is still valid for games with two opposing teams, under the assumption that teams remain static, e.g. soccer teams during a single season. Not all multiplayer games fit those criteria, however. For instance, tennis doubles, online video games or various card games are all instances of games involving two teams of possibly uneven size, and whose composition may change between games.

We now propose a variant of the BT model for games between two competing teams. The outcome of one such game is described by a pair $(I,J)$ of sets of players, which represent the two opposing teams, where $I$ denotes the winning team and $J$ the losing team. We can mathematically describe a set of games as a \textit{directed hypergraph} \cite{galloDirectedHypergraphsApplications1993,ausielloDirectedHypergraphsIntroduction2017}, with each pair $(I,J)$ corresponding to a directed hyperedge from $J$ to $I$.

Following Equation \eqref{eq:bt-prob-2}, we propose the following model for the probability $P[I \leftarrow J]$ of team $I$ winning against team $J$, based on latent player strengths $s_i \in \R, i \in \{1,\dots,n\}$,
\begin{equation} \label{eq:hbt-prob-logistic}
    P[I \leftarrow J] = f\left(\textstyle\sum_{i \in I} s_i - \textstyle\sum_{j\in J} s_j\right),
\end{equation}
where $f$ is the logistic function. Setting $\pi_i = e^{s_i}$, the analogue of Equation \eqref{eq:bt-prob} is then given by
\begin{equation} \label{eq:hbt-prob}
    P[I \leftarrow J] = \frac{\prod_{i\in I} \pi_i}{\prod_{i\in I} \pi_i + \prod_{j\in J} \pi_j}.
\end{equation}
In contrast with the BT model, the probability $P[I \leftarrow J]$ is not always invariant under adding arbitrary constants to the $s_i$'s, as the sets $I$ and $J$ are not necessarily of the same size.

From there, we wish to infer parameters $\{\pi_i\}_{i=1}^n$ from a set of recorded games $\mathcal{E} = \{(I_i,J_i)\}_{i=1}^m$, along with weights $W_{IJ}$ for each $(I,J) \in \mathcal{E}$. The log-likelihood is then given by
\begin{equation} \label{eq:hbt-log-likelihood}
    \log P(\mathcal{E}|\pi) = \sum_{(I,J)\in \mathcal{E}} W_{IJ} \left(\sum_{i \in I} \log \pi_i\right) - \sum_{(I,J) \in \mathcal{E}} W_{IJ} \log \left( \prod_{i\in I} \pi_i + \prod_{j\in J} \pi_j\right)
\end{equation}

From there, taking the gradient with respect to $\pi = [\pi_1,\dots,\pi_n]^\top$, writing the first order optimality condition and performing similar manipulations as in \cite{Newman2023}, we find the following fixed point equation, which we use as our update rule,

\begin{equation} \label{eq:hbt-update}
    \pi_k = \frac{\sum_{\substack{(I,J)\in\mathcal{E}\\k \in I}} W_{IJ} P[J \leftarrow I]}{\sum_{\substack{(I,J)\in\mathcal{E}\\k\in J}} W_{IJ} P[J \leftarrow I]} \pi_k.
\end{equation}
We provide the detailed derivation of this formula, as well as pseudocode for the final algorithm in Appendix \ref{sec:derivation}. It can also be checked that this equation reduces to \eqref{eq:bt:newman} for two player games.

In formula \eqref{eq:hbt-update}, we implicitly make the assumption that the sets $I$ and $J$ are disjoint. This is a reasonable assumption in most cases, e.g. sports, where a player can only be part of a single team at a time. Nonetheless, we derive in Appendix \ref{sec:nondisjoint} a version of our model where $I$ and $J$ are allowed to be non-disjoint and discuss potential applications of that model.

\section{Comparison with other works} \label{sec:other-works}

\subsection{Generalized Bradley-Terry Model}

The work of Huang et al. \cite{Huang2004} was, to our knowledge, the first to generalize the Bradley-Terry model to infer individual skills from paired group comparisons with a model they named Generalized Bradley-Terry model (GBT). Their approach differs crucially from ours, in that they take a different starting ansatz, and they adapt the Zermelo iteration to their model, as their work predates \cite{Newman2023}. More precisely, the authors of \cite{Huang2004} assume that the probability $P_{\operatorname{GBT}} [I \leftarrow J]$ is given by
\begin{equation} \label{eq:gbt-prob}
    P_{\operatorname{GBT}} [I \leftarrow J] = \frac{\sum_{i \in I} \pi_i}{\sum_{i\in I} \pi_i + \sum_{j \in J} \pi_j} = \frac{\pi_I}{\pi_I + \pi_J},
\end{equation}
where $\pi_S = \sum_{i \in S} \pi_i$, for some arbitrary $S\subset \{1,\dots, n\}$. This differs from Equation \eqref{eq:hbt-prob} in that products of the $\pi_i$'s are replaced by sums. As a consequence, there is no equivalent to Equation \eqref{eq:hbt-prob-logistic} for this model. Unlike our approach, Equation \eqref{eq:gbt-prob} has the advantage of remaining invariant under arbitrary scalings of the $\pi_i$'s. The two approaches can therefore be seen as complementary, with different assumptions and advantages.

Huang et al. propose the following update rule for their model, which they derive by looking for a descent direction of the negative log-likelihood,
\begin{equation} \label{eq:gbt-update}
    \pi_k = \frac{\sum_{\substack{(I,J)\in\mathcal{E}\\k \in I}} \frac{W_{IJ}}{\pi_I} - \sum_{\substack{(I,J)\in\mathcal{E}\\k \in J}} \frac{W_{IJ}}{\pi_J}}{\sum_{\substack{(I,J)\in\mathcal{E}\\k \in I \cup J}} \frac{W_{IJ}+W_{JI}}{\pi_I + \pi_J}} \pi_k.
\end{equation}

In our experiments, we found that using this formula to infer the parameters requires many iterations to reach an acceptable tolerance. We address this problem by deriving the following update, based on a similar derivation as for our own approach.
\begin{equation} \label{eq:gbt-update-alt}
    \pi_k = \frac{\sum_{\substack{(I,J)\in \mathcal{E}\\k \in I}} W_{IJ} \frac{\pi_J}{\pi_I (\pi_I + \pi_J)}}{\sum_{\substack{(I,J)\in \mathcal{E}\\k \in J}} W_{IJ} \frac{1}{\pi_I + \pi_J}} \pi_k
\end{equation}

As for Equation \eqref{eq:hbt-update}, it can be checked that Equations \eqref{eq:gbt-update} and \eqref{eq:gbt-update-alt} reduce respectively to the Zermelo update (Eq. \eqref{eq:bt-zermelo}) and Newman update (Eq. \eqref{eq:bt:newman}) for two player games. Our new update \eqref{eq:gbt-update-alt} is therefore the equivalent of the Newman update for the GBT model.

\subsection{Plackett-Luce Model}

The Plackett-Luce (PL) model \cite{Luce1959,plackettAnalysisPermutations1975} is a well-known generalization of the classical Bradley-Terry model to multiplayer games. Unlike our setting, the PL model specifically describes games whose outcome is an ordering of all the players involved, such as a race where competitors finish in a given position.

The PL model can thus be described as a higher-order version of the BT model, but in a different sense than the one we study here. Specifically, \cite{yeungEfficientInferenceRankings2025} described outcomes of the PL model as regular (undirected) hyper-edges equipped with an ordering and derived a Newman style update for the PL model.

\section{Experiments} \label{sec:experiments}

\subsection{Synthetic Data}

We start by evaluating the performance of our method on synthetic datasets. We generate random directed hypergraphs using on our model in Equation \eqref{eq:hbt-prob}, with player strengths sampled from a standard Gaussian distribution. Each hyper-edge consists of four players sampled without replacement from a pool of $n$ players and split into two teams of two with probability $0.9$ or into one team of one and a team of three with probability $0.1$.

We use Equation \eqref{eq:hbt-update} to estimate the parameters $s_i = \log \pi_i$ from these randomly generated hypergraphs. In addition, we compare it with the parameters estimated by the GBT model using Equation \eqref{eq:gbt-update-alt}, as well as the parameters estimated by the BT model using Equation \eqref{eq:bt:newman} applied to the data by considering each directed hyper-edge $(I,J)$ as a win of every $i \in I$ against every $j\in J$. 

\begin{figure}
    \includegraphics[width=0.45\textwidth]{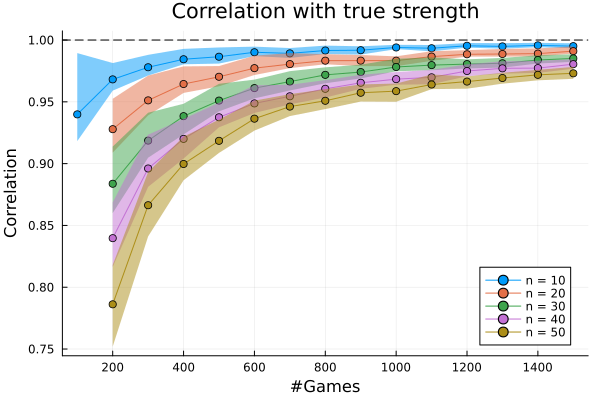}
    \includegraphics[width=0.45\textwidth]{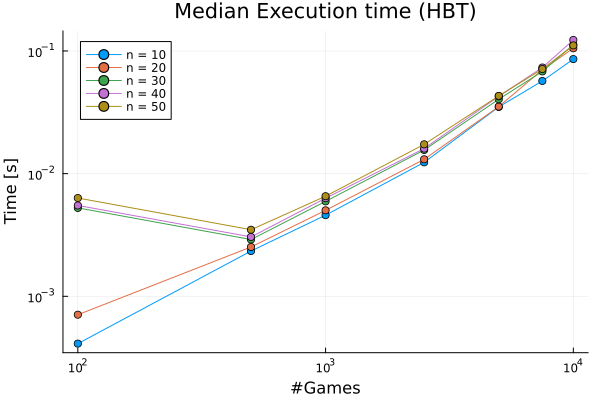}
    \caption{Performance of our method on synthetic data. \textbf{Left:} Pearson Correlation between fitted strengths $\hat{s}_i$ and true strengths $s_i$. We plot the median over $100$ replications of the correlation for multiple numbers $n$ of players and $m$ of hyper-edges. The colored bands around the points indicate the upper and lower quartiles. \textbf{Right:} Median execution time of the algorithm for multiple values of $n$ and $m$. Medians are computed over $100$ replications.}
    \label{fig:experiments}
\end{figure}

We report on the left panel of Figure \ref{fig:experiments}, the correlation between the player strengths fitted by the algorithm and the true strengths used to generate the data. We find that the estimated values are strongly correlated with the real values, and that the correlation increases with the number of edges. The fitted parameters from the GBT and BT models show similar degrees of correlation with the fitted HBT parameters.

We also examine the efficiency of our method as a function of the number of players and hyper-edges in the right panel \ref{fig:experiments}, and find that it is dominated by the number of hyper-edges, as the time to estimate the parameters varies linearly with the number of hyper-edges, with a comparably small increase with the number of players. In comparison, we find that the time to estimate parameters for the BT model is essentially constant with respect to the number of hyper-edges. This is explained by the fact the inference algorithm for the BT model takes as input a matrix that aggregates the information of all hyper-edges, and takes few iterations to converge.

\subsection{Games of Whist}

The game of Whist is a trick-taking card game originating in 18th-century Britain, with many regional variants, and similar to other games such as Bridge, Spades, Euchre, Klaverjas, or Belote. In addition to being the authors' favorite lunchtime activity, it serves as a perfect application for our model, as players in a Whist game (usually four players) are split into two teams of two players, or one player against three\footnote{We provide a detailed explanation of our specific variant of the rules in Appendix \ref{sec:whist}}.

For the purpose of this work, we have recorded a dataset of over 300 games of Whist within a total pool of $23$ players. Removing players with too few games, we end up with a total of $300$ games between $17$ players. We use the three different methods described in the previous section to establish rankings between players based on fitted strength values, and compare those against the ranking based on the ratio of the number of games won over the total number of games played\footnote{We avoid stating precise numbers for fear of retribution.}. We remark that all of these quantities are strongly correlated with one another, as the Pearson correlation between any pair of them is greater than $0.9$.

The rankings are reported in Table \ref{tab:rankings}, where we observe that while no two methods produce exactly the same ranking, all rankings broadly agree in a looser sense. Player A is the only player whose rank is always the same, followed by some permutation of players B, C and I, and the top 8 players are nearly all the same across rankings. Players E and G are consistently at the bottom of the rankings. Surprisingly, some players such as player K who rank low based on win rate are ranked higher by the other rankings. Finally, we note that all rankings appear to be sensitive to the number of games played by each player, as players with small number of games tend to land on the extreme ends of the rankings (this is the case for players B, E and G who each have played less than twenty games).

\begin{table}
    \centering
    \begin{tabular}{c|c|c|c|c}
        Rank & Win Rate & HBT & GBT & BT \\ \hline
        1    & A        & A   & A   & A \\
        2    & I        & C   & B   & B \\
        3    & C        & B   & I   & I \\
        4    & F        & I   & P   & C \\
        5    & B        & F   & C   & F \\
        6    & P        & P   & F   & P \\
        7    & H        & H   & H   & Q \\
        8    & Q        & D   & Q   & H \\
        9    & O        & K   & D   & L \\
        10   & D        & Q   & N   & J \\
        11   & J        & M   & L   & K \\
        12   & N        & J   & M   & N \\
        13   & L        & L   & K   & D \\
        14   & M        & N   & O   & O \\
        15   & K        & O   & J   & M \\
        16   & G        & G   & E   & E \\
        17   & E        & E   & G   & G
    \end{tabular}
    \caption{Ranking of the authors based on the various methods. Authors labelled A to Q in alphabetical order.}
    \label{tab:rankings}
\end{table}

\section{Conclusion}

We have introduced a novel statistical model of games with two opposing teams of players along with an efficient method for inferring individual player strengths from such games. In addition, we derived a similarly efficient method for the prior model of \cite{Huang2004}. We found in our experiments that the parameters estimated by all the models we considered (BT, HBT, GBT) are strongly correlated with each other.

This being said, there are many important questions which are left unaddressed by this work. In particular, we have not discussed the convergence of our method to a global minimum, which is well-studied for the classical BT model \cite{zermeloBerechnungTurnierErgebnisseAls1929,Newman2023}. We have also kept our experiments to simple settings, and have not explored the behavior of the various methods as the number of players in a team varies.

We plan to address these points in upcoming, less frivolous work.

\bibliographystyle{unsrt} 
\bibliography{sources}

\newpage
\appendix
\section{Full derivations and pseudocode} \label{sec:derivation}

\subsection{Classical Bradley-Terry model}

Recall from Equation \eqref{eq:bt-loglikelihood} that the log-likelihood is given by
$$ \log P(W | \pi) = \sum_{i < j} W_{ij} \log \left(\frac{\pi_i}{\pi_i + \pi_j}\right) + W_{ji} \log \left( \frac{\pi_j}{\pi_i + \pi_j} \right). $$
Taking the partial derivative with respect to $\pi_i$, we find
\begin{equation}
    \frac{\partial}{\partial \pi_i} \log P(W|\pi) = \sum_{j} \frac{W_{ij}}{\pi_i} - \frac{W_{ij}}{\pi_i + \pi_j} - \frac{W_{ji}}{\pi_i + \pi_j}.
\end{equation}
Setting $\frac{\partial}{\partial x_i}\log P(W|\pi)=0$, Zermelo's fixed-point Equation \eqref{eq:bt-zermelo} is then obtained by grouping the last two terms together, then performing some elementary algebra to isolate $\pi_i$. Newman's equation \eqref{eq:bt:newman} is simply obtained by grouping the first two terms instead. The pseudocode is given in Algorithm \ref{alg:bt}.

\begin{algorithm}
    \caption{Newman's algorithm for inferring BT parameters}
    \label{alg:bt}
    \begin{algorithmic}
        \Function{Bradley-Terry-MLE}{$W\in \R^{n\times n}$}
        \State $\pi \gets [1,\dots,1]^\top \in \R^n$
        \While{not converged}
        \For{$i = 1,\dots,n$}
            \State $\pi_i \gets \frac{\sum_{j=1}^n W_{ij} \frac{\pi_j}{\pi_i + \pi_j}}{\sum_{j=1}^n W_{ji} \frac{1}{\pi_i + \pi_j}}$ \Comment{Update each parameter one by one}
        \EndFor
        \State \Call{normalize}{$\pi$} \Comment{Normalize parameters (optional)}
        \EndWhile
        \State \Return $\pi$
        \EndFunction
    \end{algorithmic}
\end{algorithm}

\subsection{Hypergraph Bradley-Terry model}

Starting from \eqref{eq:hbt-prob}, we note $\Pi_S = \prod_{i\in S} \pi_i$ for conciseness. The log-likelihood is then written as
\begin{equation}
    \log P(\mathcal{E}| \pi) = \sum_{(I,J)\in\mathcal{E}} W_{IJ} \sum_{i\in I} \log \pi_i - \sum_{(I,J)\in \mathcal{E}} W_{IJ} \log \left(\Pi_I + \Pi_J\right).
\end{equation}
Taking the partial derivative with respect to $\pi_k$, only the terms in the above sums in which $k$ is a member of either $I$ or $J$ are relevant. In addition, noting that if $k\in S$, $\frac{\partial \Pi_S}{\partial \pi_k} = \Pi_S / \pi_k$, we may write
\begin{align}
    \frac{\partial}{\partial \pi_k} \log P(\mathcal{E}|\pi) &= \sum_{\substack{(I,J)\in\mathcal{E}\\k\in I}} W_{IJ} \frac{1}{\pi_k} - \sum_{\substack{(I,J)\in\mathcal{E}\\k\in I}} W_{IJ} \frac{\Pi_I}{\Pi_I + \Pi_J} \frac{1}{\pi_k} - \sum_{\substack{(I,J)\in\mathcal{E}\\k\in J}} W_{IJ} \frac{\Pi_J}{\Pi_I + \Pi_J} \frac{1}{\pi_k} \\
    &= \sum_{\substack{(I,J)\in\mathcal{E}\\k\in I}} W_{IJ} \frac{1}{\pi_k} - \sum_{\substack{(I,J)\in\mathcal{E}\\k\in I}} W_{IJ} P[I\leftarrow J] \frac{1}{\pi_k} - \sum_{\substack{(I,J)\in\mathcal{E}\\k\in J}} W_{IJ} P[J\leftarrow I] \frac{1}{\pi_k} \\
    &= \frac{1}{\pi_k} \sum_{\substack{(I,J)\in\mathcal{E}\\k \in I}} W_{IJ} P[J\leftarrow I] - \frac{1}{\pi_k} \sum_{\substack{(I,J)\in\mathcal{E}\\k\in J}} W_{IJ} P[J\leftarrow I],
\end{align}
where in the last line, we have grouped the first two sums and used the fact that $1-P[I\leftarrow J] = P[J\leftarrow I]$. Equation \eqref{eq:hbt-update} then follows straightforwardly. Unlike the classical BT model, we found that is unnecessary, and even harmful to normalize $\pi$ in Algorithm \ref{alg:hbt}. This is due to Equation \eqref{eq:hbt-prob} no longer being invariant under rescaling $\pi$.

\begin{algorithm}
    \caption{Inference of HBT parameters}
    \label{alg:hbt}
    \begin{algorithmic}
        \Function{Hypergraph-Bradley-Terry-MLE}{$\mathcal{E}$, $W$}
        \State $\pi \gets [1,\dots,1]^\top \in \R^n$
        \While{not converged}
        \For{$k = 1,\dots,n$}
            \State $\pi_k \gets \frac{\sum_{\substack{(I,J)\in\mathcal{E}\\k\in I}} W_{IJ} P[J\leftarrow I]}{\sum_{\substack{(I,J)\in \mathcal{E}\\k\in J}} W_{IJ} P[J\leftarrow I]} \pi_k$ \Comment{Update each parameter one by one}
        \EndFor
        \EndWhile
        \State \Return $\pi$
        \EndFunction
    \end{algorithmic}
\end{algorithm}

\subsection{Updated equation for the Generalized Bradley-Terry model}

We derive our updated equation for the Generalized Bradley-Terry model (Eq. \eqref{eq:gbt-prob}), starting from the log-likelihood, and noting $\pi_S = \sum_{i\in S} \pi_i$, for $S \subset \{1,\dots,n\}$,
\begin{equation}
    \log P_{\operatorname{GBT}} (\mathcal{E}| \pi) = \sum_{(I,J)\in \mathcal{E}} W_{IJ} \log \pi_I - \sum_{(I,J)\in \mathcal{E}} W_{IJ} \log (\pi_I + \pi_J).
\end{equation}
Elementary calculations provide us with the following expressions for the gradient entries,
\begin{align}
    \frac{\partial}{\partial \pi_k} \log P_{\operatorname{GBT}}(\mathcal{E} | \pi) &= \sum_{\substack{(I,J)\in \mathcal{E}\\ k \in I}} W_{IJ} \frac{1}{\pi_I} - \sum_{\substack{(I,J)\in \mathcal{E}\\k\in I}} W_{IJ} \frac{1}{\pi_I + \pi_J} - \sum_{\substack{(I,J)\in \mathcal{E}\\ k\in J}} W_{IJ} \frac{1}{\pi_I + \pi_J} \\
    &= \sum_{\substack{(I,J)\in \mathcal{E}\\k \in I}} W_{IJ} \frac{\pi_J}{\pi_I (\pi_I + \pi_J)} - \sum_{\substack{(I,J)\in \mathcal{E}\\k\in J}} W_{IJ} \frac{1}{\pi_I + \pi_J} \\
    &= \sum_{\substack{(I,J)\in \mathcal{E}\\k \in I}} W_{IJ} \frac{1}{\pi_I} P_{\operatorname{GBT}} [J\leftarrow I] - \sum_{\substack{(I,J)\in \mathcal{E}\\k\in J}} W_{IJ} \frac{1}{\pi_J} P_{\operatorname{GBT}} [J \leftarrow I].
\end{align}
These equations do not contain $\pi_k$ directly. In order to obtain Equation \eqref{eq:gbt-update-alt}, we do the same manipulations as for our model, and then multiply both sides by $\pi_k$.

\begin{algorithm}
    \caption{Inference of GBT parameters}
    \label{alg:gbt}
    \begin{algorithmic}
        \Function{Generalized-Bradley-Terry-MLE}{$\mathcal{E}$, $W$}
        \State $\pi \gets [1,\dots,1]^\top \in \R^n$
        \While{not converged}
        \For{$k = 1,\dots,n$}
            \State $\pi_k \gets \frac{\sum_{\substack{(I,J)\in\mathcal{E}\\k\in I}} W_{IJ} \frac{\pi_J}{\pi_I (\pi_I + \pi_J)}}{\sum_{\substack{(I,J)\in \mathcal{E}\\k\in J}} W_{IJ} \frac{1}{\pi_I + \pi_J}} \pi_k$ \Comment{Update each parameter one by one}
        \EndFor
        \State \Call{normalize}{$\pi$} \Comment{Normalize parameters}
        \EndWhile
        \State \Return $\pi$
        \EndFunction
    \end{algorithmic}
\end{algorithm}

\subsection{Hypergraph Bradley-Terry with non-disjoint subsets} \label{sec:nondisjoint}

In most applications, we can assume that directed hyper-edges $(I,J)$ are composed of disjoint sets. Nonetheless, if we allow $I$ and $J$ to have a non-empty intersection, it is still possible to derive an update that takes this into account. One example of such a situation is ranking researcher based on a citation network. 

Given a set of published papers, we consider paper $A$ with author set $I$ citing paper $B$ with author set $J$ as a directed hyper-edge $(I,J)$ (we may also weight this hyper-edge by the number of times paper $B$ is cited in paper $A$, which serves as a measure of how much $A$ is influenced by $B$). It is common for authors to cite their previous work, so in this setting it appears necessary to allow for non-disjoint $I$ and $J$.

We can adapt our model to this setting by noticing that
\begin{equation}
    \frac{\partial}{\partial \pi_k} \log \left(\Pi_I + \Pi_J\right) = \begin{cases}
        \frac{1}{\pi_k} P[I \leftarrow J] & k \in I \setminus J\\
        \frac{1}{\pi_k} P[J \leftarrow I] & k \in J \setminus I\\
        \frac{1}{\pi_k} & k \in I \cap J\\
        0 & \text{else},
    \end{cases}
\end{equation}
and breaking down the sum in Equation \eqref{eq:hbt-log-likelihood} in these different cases. This gives
\begin{align*}
    \frac{\partial}{\partial \pi_k} \log P(\mathcal{E}|\pi) &= \sum_{\substack{(I,J)\in\mathcal{E}\\k\in I}} W_{IJ} \frac{1}{\pi_k} - \sum_{\substack{(I,J)\in\mathcal{E}\\k\in I\setminus J}} W_{IJ} \frac{P[I \leftarrow J]}{\pi_k} - \sum_{\substack{(I,J)\in\mathcal{E}\\k\in J\setminus I}} W_{IJ} \frac{P[J\leftarrow I]}{\pi_k} - \sum_{\substack{(I,J)\in\mathcal{E}\\k \in I \cap J}} W_{IJ} \frac{1}{\pi_k}\\
    &= \sum_{\substack{(I,J)\in\mathcal{E}\\k \in I \setminus J}} W_{IJ} \frac{1 - P[I\leftarrow J]}{\pi_k} - \sum_{\substack{(I,J)\in\mathcal{E}\\k\in J\setminus I}} W_{IJ} \frac{P[J\leftarrow I]}{\pi_k} \\
    &= \sum_{\substack{(I,J)\in\mathcal{E}\\k \in I \setminus J}} W_{IJ} \frac{P[J\leftarrow I]}{\pi_k} -  \sum_{\substack{(I,J)\in\mathcal{E}\\k\in J\setminus I}} W_{IJ} \frac{P[J\leftarrow I]}{\pi_k}.
\end{align*}
In the above equation, we see that the contribution from hyper-edges where $k \in I \cap J$ cancels out to zero. This gives us the following update rule
\begin{equation} \label{eq:hbt-update-non-disjoint}
    \pi_k = \frac{\sum_{\substack{(I,J)\in\mathcal{E}\\k \in I\setminus J}} W_{IJ} P[J \leftarrow I]}{\sum_{\substack{(I,J)\in\mathcal{E}\\k\in J\setminus I}} W_{IJ} P[J \leftarrow I]} \pi_k.
\end{equation}

The same derivation can be performed for the GBT model, which produces the following update rule,
\begin{equation}
    \pi_k = \frac{\sum_{\substack{(I,J)\in\mathcal{E}\\k\in I\setminus J}} W_{IJ} \frac{\pi_J}{\pi_I (\pi_I + \pi_J)} + \sum_{\substack{(I,J)\in\mathcal{E}\\k \in I\cap J}} W_{IJ} \frac{\pi_J - \pi_I}{\pi_I (\pi_I + \pi_J)}}{\sum_{\substack{(I,J)\in\mathcal{E}\\k\in J\setminus I}} W_{IJ} \frac{1}{\pi_I + \pi_J}} \pi_k.
\end{equation} 
In contrast with Equation \eqref{eq:hbt-update-non-disjoint}, we see that the contribution of hyper-edges with non-disjoint $I$ and $J$ does not vanish.

\section{The Game of (Namur) Whist} \label{sec:whist}

There are multiple related games going by the name "Whist", none of which exactly match the specific rule set used to create the dataset in our main experiment\footnote{It is most closely related to the "Whist à la couleur" variant, which is the version of the game popular in Belgium.}. We therefore provide a description of those rules for reference\footnote{We encourage the reader who read this far to try it with friends.}.

\subsection{Setup}

The game of Whist (or, to disambiguate, of "Namur Whist") is played by four players using a standard deck of 52 playing cards. The players (hereafter $A$, $B$, $C$ and $D$) sit in clockwise order at a table and the dealer ($A$), starts by shuffling the cards, then asking the player to their right ($D$) to cut the deck. They then distribute cards in clockwise order, starting with the player to their left ($B$), with a first round of four cards per player, then five cards, then four cards again.

\subsection{Announcements}

Once all the cards are dealt, each player must check their hand to see if it contains three aces or more and announce it if it does. In that case, the game is in a state of "Trou" (hole) and special rules apply (see below).

If no hole was announced, player make bids to decide the game's trump. Starting from player $B$ in clockwise order, each player may take one of the following actions
\begin{enumerate}
    \item Announce a suit that has not been announced by another player.
    \item Follow a suit announced by a previous player if it has not already been followed. 
    \item Pass, in which case the player no longer participates in the bidding.
    \item Announce a special hand (detailed below).
\end{enumerate}
In addition, player $B$ may choose to declare "First", in which case they do not announce a suit, but may follow another player after all other players have made their announcement. After that, any single player that hasn't passed may choose to either play alone, follow another player who announced a suit higher than them (based on the order $\spadesuit < \clubsuit < \diamondsuit < \heartsuit$) or pass.

\subsection{Bidding}

After all announcements have been made, each player is either alone, in a team with another player or out of the bidding. Bidding is done by declaring how many rounds one expects to win. A single player bids directly, whereas in a team, the player who followed is the one making the bids, with no allowed input from the other player.

Starting with the team with the lowest-ranked suit (based on the order $\spadesuit < \clubsuit < \diamondsuit < \heartsuit$), each team declares how many rounds they will have to win, or folds. A single player starts at six rounds, while a two-player team starts at eight. Once all teams have either declared or folded, the next round of bidding begins, with all bids increased by one. If the player in a team of two folds, the remaining player may decide to go alone, with a bid decreased by two. This continues until all but one team has folded. The remaining team's suit will be the trump for the game, and they need to win at least as many rounds as their bid in order to win.

\subsection{Play}

The game proper consists in thirteen rounds in which each player throws a card on the table. The first player ($B$ in the first round, then the winner of the last round) throws the first card (whose suit becomes the round's suit), then the other players each throw a card in clockwise order. The round is won by the player who played the highest card (in standard order, with $2$ being the lowest card to Ace being the highest) in the round's suit, or if any trumps are played, by the player who played the highest trump.

After all rounds have been played, if the announcing players have won at least as many rounds as their bid, they win the game. Otherwise, the remaining players win.

\subsection{Special hands}

There are various special hands, which are ordered by priority based on their difficulty. We list them below from lowest to highest.

\begin{description}
    \item[Petite Misère] A player picks a card in their hand and places it face down on the table. Unless contested by a hand with higher priority, the other players also remove a card from their hand. The game then starts normally, except there is no trump, and the player who announced the misère needs to win no rounds in order to win. A second player may also decide to play a Petite Misère, in which case both players play separately.
    \item[Picolissimo] A single player has to win exactly two hands. There is no trump.
    \item[Piccolo] A single player has to win exactly one round. There is no trump.
    \item[Abondance] The player announces "Abondance" with a suit. Unless contested by a higher hand, they play alone and have to win at least nine rounds.
    \item[Trou] If one player has three aces in their hand, they must declare it. Unless contested by a Grande Misère (in which case, it is a Grande Misère sur Trou), the player with the remaining ace, without communicating with the other players, plays the first card of the first round, whose suit becomes the trump. If the first card played is an ace, the team with the aces must win at least nine rounds. If the first card is not an ace, the player with the three aces has to play the corresponding ace and the team must win at least ten rounds.
    \item[Grande Misère] Same as a Petite Misère, except no cards are discarded.
    \item[Grande Misère sur table] Same as a Grande Misère, except the announcing player must lay all their cards face visible on the table.
    \item[Trou Royal] If a player has four aces in their hand, they must announce a "Trou Royal". The trump is Hearts and the player with the highest Hearts card beside the announcer begins the first round as in a normal Trou.
    \item[Solo Chelem] A single player has to win all thirteen rounds. There is no trump. 
    \item[Chasse c\oe ur - Chasse dame] In the event that no announcements are made in the first round of bidding, a special rule applies. The game proceeds with no trumps and no teams, and a player wins if they either won no rounds containing a queen card, or if they have won all rounds containing queens.
\end{description}

\end{document}